\documentclass[11pt]{article}
\pdfoutput=1
\usepackage{verbatim,url,enumerate,color,paralist}
\usepackage{amsmath,amsfonts}
\usepackage{epsfig,amssymb,amstext,xspace,theorem}
\usepackage{algorithm}
\usepackage{algorithmicx,algpseudocode}
\usepackage{pifont}

\usepackage{caption}
\usepackage{slashbox}
\usepackage{graphicx}
\usepackage{tikz}

\newcommand{\qed}{\hspace*{\fill}$\Box$}

\newcommand{\bs}{\backslash}
\newcommand{\IGNORE}[1]{}

\newtheorem{theorem}{Theorem}[section]
\newtheorem{lemma}[theorem]{Lemma}
\newtheorem{corollary}[theorem]{Corollary}

\newtheorem{remark}[theorem]{Remark}

\newenvironment{proof}[1][Proof. ]{\noindent {\bf #1 }}{\qed}
\begin{document}

\title {On the Metric $s$-$t$ Path Traveling Salesman Problem}
\author{
Zhihan Gao\thanks{
        (z9gao@uwaterloo.ca)
	Dept.\ of Comb.\ \& Opt.,
        University of Waterloo, Waterloo, Ontario N2L3G1, Canada.
	}
}

\date{}
\maketitle

\begin{abstract}
We study the metric $s$-$t$ path Traveling Salesman Problem (TSP).
[An, Kleinberg, and Shmoys, STOC 2012] improved on
the long standing $\frac{5}{3}$-approximation factor and
presented an algorithm that achieves an approximation factor of
$\frac{1+\sqrt{5}}{2}\approx1.61803$.
Later [Seb\H{o}, IPCO 2013] further improved the approximation factor
to $\frac{8}{5}$.
We present a simple, self-contained analysis that
unifies both results;
our main contribution is a \emph{unified correction vector}.
Additionally, we compare two different linear programming (LP) relaxations of
the $s$-$t$ path TSP, namely,
the path version of the Held-Karp LP relaxation for TSP
and a weaker LP relaxation, and we show that both LPs have
the same (fractional) optimal value. Also, we show that the minimum cost of integral solutions of the
two LPs are within a factor of $\frac{3}{2}$ of each other.
Furthermore, we prove that a half-integral solution of the stronger LP-relaxation of
cost $c$ can be rounded to an integral solution of cost at most $\frac{3}{2}c$.
%
Finally, we give an instance that presents obstructions to
two natural methods that aim for an approximation factor of $\frac{3}{2}$.
\end{abstract}


\section{Introduction}

The metric Traveling Salesman Problem (TSP) is a celebrated
problem in Combinatorial Optimization, see \cite[Chapter 58]{Sch03}, \cite{BB08}. One important variant of TSP
is the (metric) $s$-$t$ path TSP. Let $G$ be a complete graph $G$ with
nonnegative metric edge costs $c$, i.e., $c$ satisfies the triangle inequality.
Given two fixed vertices $s, t$ in $G$, the
\emph{$s$-$t$ path TSP} is to find a minimum-cost Hamiltonian path
from $s$ to $t$ in $G$.

Hoogeveen \cite{hoogeveen91} gave an $s$-$t$ path TSP variant of
Christofides' approximation algorithm for the TSP \cite{christofides76}, and obtained an approximation factor
of $\frac{5}{3}$. There was no improvement in this approximation factor
for over two decades until An, Kleinberg, and Shmoys \cite{AKS12}
improved the approximation factor to $\frac{1+\sqrt{5}}{2}\approx1.61803$.
One of the key new contributions of \cite{AKS12} is to
design and analyse a randomized version of Christofides' algorithm.
The analysis introduced the notion of a correction vector for
the $s$-$t$ path TSP.
Most recently, Seb\H{o} \cite{sebo13} further improved the
analysis and obtained a better approximation factor of $\frac{8}{5}$.
\cite{sebo13} introduced a correction vector different from that
of \cite{AKS12}, and this is one reason why the analysis in \cite{sebo13}
gives a better approximation factor. Informally speaking,
a better correction vector provides a better approximation factor.
In this paper, we give a unified presentation of the results from both
\cite{AKS12} and \cite{sebo13} by introducing a new correction vector
that we call the \emph{unified correction vector}.
Our correction vector is simple and it leads to short derivations of
the approximation factors of both \cite{AKS12} and \cite{sebo13}. The difference between our correction vector and
the previous ones is that it assigns the value one to the minimum-cost
edge in each so-called $\tau$-narrow cut, whereas the correction vectors used in \cite{AKS12} and \cite{sebo13} are fractional
on each $\tau$-narrow cut. We mention that Vygen's \cite{vygen13} comprehensive recent survey discusses
the common points of the analysis of \cite{AKS12} and \cite{sebo13},
and the survey sketches short proofs of both approximation factors;
however, \cite{vygen13} uses the same correction vectors as \cite{AKS12} and \cite{sebo13}.
%

An et al. \cite{AKS12} and Seb\H{o} \cite{sebo13} use two
different LP relaxations of the $s$-$t$ path TSP in their algorithms.
\cite{AKS12} uses the path version of the Held-Karp LP relaxation for TSP,
whereas \cite{sebo13} uses a weaker LP relaxation. This motivates a comparison of these two LP relaxations. We mention that Seb\H{o} proves an approximation factor of $\frac{8}{5}$ for a more
general problem, namely, the \emph{connected T-join problem}, and the LP in
his paper is a relaxation of this problem.
We show that both LPs for the $s$-$t$ path TSP have the same (fractional) optimal value.
Also, we show that the minimum cost of integral solutions of the
two LPs are within a factor of $\frac{3}{2}$ of each other; moreover,
we present an example to show that the factor of $\frac{3}{2}$ is tight.
We prove this result by showing that
a half-integral solution of the stronger LP-relaxation of cost $c$
can be rounded to an integral solution of cost at most $\frac{3}{2}c$.


For the $s$-$t$ path TSP, it is known that the integrality ratio
of the path version of the Held-Karp LP relaxation has a lower bound
of $\frac{3}{2}$. All of the algorithms mentioned above are LP-based.
This leads to the best known upper bound $\frac{8}{5}$ on the integrality
ratio of the LP relaxation. A natural open question is to close this
gap by designing an LP-based $\frac{3}{2}$-approximation algorithm
for the $s$-$t$ path TSP. Given a connected graph $H$ with unit
edge costs and two fixed vertices $s$ and $t$, the \emph{$s$-$t$
path graph-TSP} is to find a minimum-cost Hamiltonian path from $s$
to $t$ in the metric completion of $H$. For this critical special
case of the $s$-$t$ path TSP, the integrality ratio of the corresponding
LP relaxation has been resolved already. The first
$\frac{3}{2}$-approximation algorithm was given by Seb\H{o} and Vygen
\cite{SV12} using ear decompositions.  Gao \cite{Gao13} designed
another, conceptually simpler, LP-based
$\frac{3}{2}$-approximation algorithm.
The analysis of the $\frac{3}{2}$-approximation factor of \cite{Gao13}
uses the graphic property only for one point: to guarantee that the
cost of a special spanning tree constructed in the algorithm is at
most the optimum of the LP relaxation. A natural question is whether we can
extend this graphic LP-based approximation algorithm and analysis
to the general metric case. Unfortunately, we present an instance
that shows that that is not possible.  Moreover, our instance also illustrates that probabilistic methods
are relevant for the analysis of improved LP-based approximation
algorithms.  This instance may shed some
light on how to design a better approximation algorithm for the
$s$-$t$ path TSP.

The paper is organized as follows. Section~\ref{sec:pre} has some
notation and basic results. Section~\ref{sec:ucv} presents our
unified correction vector. Section~\ref{sec:LPs} shows the relationship
of two different LP relaxations of the $s$-$t$ path TSP.
Section~\ref{sec:BI} discusses an instance that points to some
of the obstructions for obtaining better approximation factors.


\section{Preliminaries}\label{sec:pre}

Let $G=(V, E)$ be a complete graph.
Let $s,t$ be two fixed vertices in $G$.
%
%
We call a nonempty, proper subset of vertices $S$ a \emph{cut};
thus, $\emptyset\subsetneq S\subsetneq V$.
In particular, if $|S \cap \{s, t\}|=1$,
then we call $S$ an \emph{$s$-$t$ cut}.
For $S\subseteq{V}$, let $\delta(S)$ denote the set of edges
that have one end in $S$, thus,
$\delta(S)=\{(u,v) \in E: u\in S,  v\notin S\}$.
If $S=\{v\}$, then we use $\delta(v)$ instead of $\delta(\{v\})$.
Let $E(S)$ denote the set of edges induced by $S$, thus,
$E(S) = \{(u,v)\in E: u, v \in S\}$.
For any two sets $A$ and $B$,
we use $A\backslash B$ to denote $\{a\in A: a\notin B\}$. For a vector $x\in \mathbb{R}^{A}$, we define $x(D)=\sum_{e\in D}x(e)$
for any subset $D$ of $A$. When there is no risk of confusion, we will use the same notation $H$ for a subgraph $H$ and its edge set $E(H)$.

For any probabilistic event $A$,
we use $\Pr(A)$ to denote the probability of occurrence of $A$.
For a random variable $R$,
the expectation of $R$ is denoted by $\mathbb{E}(R)$.


\subsection{Linear programs}

The path version of the Held-Karp relaxation for
the $s$-$t$ path TSP is defined as follows:

$
     \begin{aligned}
     \hbox{({\bf L.P.1})}\quad  {\rm minimize}: & \sum_{e \in E} c_ex_e &&  \\
{\rm subject~to}: & \ x(\delta(s))= x(\delta(t))= 1 &&  \\
& \ x(\delta(v))=2   && \forall \ v\neq s, t   \\
& \ x(\delta(S))\geq 1 && \forall \ s\text{-}t \mbox{ cut } S   \\
& \ x(\delta(S)) \geq  2  && \forall \ \emptyset \subsetneq S \subsetneq V, |S \cap \{s, t\}| \mbox{ even }   \\
& \ 1 \geq x_e  \geq   0  && \forall \ e \in E
     \end{aligned}
$

The spanning tree polytope is shown as follows:

$
     \begin{aligned}
     \hbox{({\bf L.P.2})}\quad  {\rm minimize} : &\ \sum_{e \in E} c_ex_e && \\
{\rm subject~to} : & \ x(E)  = |V|-1  && \\
& \ x(E(S))  \leq  |S|-1 && \forall  \emptyset\subsetneq S\subsetneq V  \\
& \ x_e  \geq  0 && \forall e \in E
     \end{aligned}
$

\begin{lemma}
Every solution $x$ of (L.P.1) lies in the spanning tree polytope (L.P.2).
\end{lemma}
\begin{proof}
By the degree constraint for each vertex in (L.P.1), we have $x(E)  = |V|-1$. Now consider the second set of constraint in (L.P.2). If $|S \cap \{s, t\}|$ is even, by the degree and cut constraints in (L.P.1), $x(E(S)) = \frac{\sum_{v\in S} x(\delta(v)) - x(\delta(S))}{2} \leq \frac{2|S|-2}{2}= |S|-1$. Otherwise, $|S \cap \{s, t\}|=1$. Similarly, $x(E(S)) = \frac{\sum_{v\in S} x(\delta(v)) - x(\delta(S))}{2} \leq \frac{(2|S|-1)-1}{2}=|S|-1$. This completes the proof.
\end{proof}


\subsection{$T$-joins}

Let $T$ be a nonempty subset of $V$ with $|T|$ even.
For $F\subseteq E$, if the set of odd degree vertices of
the graph $(V, F)$ is $T$, then we call $F$ a \emph{$T$-join}.
For any $\emptyset \subsetneq S\subseteq V$, if $|S\cap T|$ is odd (even),
then we call $S$ a \emph{$T$-odd cut} (\emph{$T$-even cut}).
The following LP formulates the problem of
finding a $T$-join of minimum cost:
\medskip

$
\begin{array}{rrcll}
\hbox{({\bf L.P.3})}\quad  {\rm minimize} : & \sum_{e \in E} c_ex_e & & \\
{\rm subject~to} : & x(\delta(S)) & \geq & 1 & \forall \mbox{ $T$-odd } S \\
& x_e & \geq & 0 & \forall e \in E
\end{array}
$
\begin{lemma}\cite{EJ01}\label{lem:Tjoin}
The optimal value of (L.P.3) is the same as the minimum cost of a $T$-join.
\end{lemma}

Let $K$ be a spanning tree. \emph{The set of wrong degree vertices} of $K$ is defined as $\{v\in \{s, t\}: |\delta(v)\cap K| \mbox{ even }\} \cup
\{v\in V\backslash \{s, t\} : |\delta(v)\cap K| \mbox{ odd }\}$.
\begin{lemma}\cite{AKS12}\label{lem:simpleParity}
Let $T$ be the set of wrong degree vertices of a spanning tree $K$. Let $S$ be an $s$-$t$ cut. If $S$ is $T$-odd, then $|\delta(S)\cap K|$ is even.
\end{lemma}

The proof can be also found in \cite[Lemma 2.1]{CFG12}. But for the sake of completeness, we present a proof here.

\vspace{3mm}

\begin{proof}
Since $\sum_{v\in S}|\delta(v)\cap K|=2|E(S)\cap K| + |\delta(S)\cap K|$, we have $|\delta(S)\cap K|$ has the same parity as $\sum_{v\in S}|\delta(v)\cap K|$. Without loss of the generality, we assume $s\in S, t\notin S$. By the definition of $T$, we know that $(S\backslash \{s\})\cap T$ is the set of vertices $v$ in $S\backslash \{s\}$ such that $|\delta(v)\cap K|$ is odd. If $|\delta(s)\cap K|$ is odd, then $s\notin T$. In this case, since $S$ is $T$-odd, $|(S\backslash \{s\})\cap T|$ is odd.  Hence, we have an even number of vertices $v$ in $S$ such that $|\delta(v)\cap K|$ is odd, which implies that $\sum_{v\in S}|\delta(v)\cap K|$ is even. Otherwise, $|\delta(s)\cap K|$ is even. Then, $s\in T$. This implies that $|(S\backslash \{s\})\cap T|$ is even. Similarly, $\sum_{v\in S}|\delta(v)\cap K|$ is even.
\end{proof}

\subsection{Polyhedra and convex decomposition}

Let
\[
\mathcal{P}:=\{x: Ax\leq b\} \quad \text{where } A\in \mathbb{R}^{m\times n}, b\in \mathbb{R}^{m}.
\]
Let $x'$ be a feasible solution of $\mathcal{P}$. For a constraint ${a_i}^{^\intercal}x\leq b_i$ in $\mathcal{P}$, we say $x'$ is \emph{tight} at this constraint if ${a_i}^{^\intercal}x'= b_i$. Let $x_1, x_2$ be two distinct feasible solutions of $\mathcal{P}$. If there exists a $0<\lambda<1$ and $y\in \mathcal{P}$
such that $\lambda x_1+(1-\lambda) y=x_2$, we say
$x_1$ is \emph{in some convex decomposition} of $x_2$ in $\mathcal{P}$.

From the geometry of polyhedra, we have the following characterization of the convex
decompositions.

\begin{lemma}\label{lem:decomposition}
The solution $x_1$ is in some convex decomposition of $x_2$ in $\mathcal{P}$ if and only if $x_1$ is tight at the constraints of $\mathcal{P}$ where $x_2$ is tight.
\end{lemma}

\subsection{Christofides' algorithm for $s$-$t$ path TSP}
Hoogeveen \cite{hoogeveen91} gave a variant of Christofides' algorithm to achieve the first approximation
factor of $\frac{5}{3}$ for the $s$-$t$ path TSP.\\

\noindent \textbf{Christofides' algorithm for $s$-$t$ path TSP}\\
\noindent Compute a minimum-cost spanning tree $J^*$. Let $T$ be the set of wrong degree vertices of $J^*$.
Find a minimum-cost $T$-join $F^*$. Then, the union $J^*\dot{\cup}F^*$ of $J^*$ and $F^*$ (that keeps the duplicated edges) forms a
connected graph that has even degree at all nodes except $s$ and $t$. One can then take the Eulerian traversal that starts
at $s$ and ends at $t$, and shortcut it, to obtain
an $s$-$t$ path visiting all vertices of no greater cost.
\begin{theorem}\cite{hoogeveen91}
Christofides' algorithm for $s$-$t$ path TSP achieves an approximation factor of $\frac{5}{3}$.
\end{theorem}

For the sake of completeness, we present a nice proof from Seb\H{o} and Vygen \cite{SV12}.

\vspace{3mm}

\begin{proof}
Let $P^*$ be an optimal solution of $s$-$t$ path TSP. Let $T, J^*, F^*$ be as in the algorithm. Let $R$ be the $s$-$t$ path in $J^*$
 and $F_{P^*}$ be the $T$-join in $P^*$. Since $P^*$ is a spanning tree,
 we know $c(J^*)\leq c(P^*)$. So, we only need to prove $c(F^*)\leq \frac{2}{3} c(P^*)$.
 This follows from the fact that $J^*\dot{\cup}P^*$ can be partitioned into three $T$-joins:
 one is $J^*\backslash R$, one is $F_{P^*}$,
 and one is the union of $R$ and $P^*\backslash F_{P^*}$. One can check that each of these edge sets is a $T$-join
 by using the fact that $T$ is the set of wrong degree vertices of $J^*$.
  Then, $3c(F^*)\leq c(J^*)+c(P^*)\leq 2c(P^*)$. This completes the proof.
\end{proof}

\section{Unified correction vector}\label{sec:ucv}

An et al. \cite{AKS12} designed a randomized Christofides' algorithm for the $s$-$t$ path~TSP, and
they proved an approximation factor of
$\frac{1+\sqrt{5}}{2}$ by analysing this algorithm.
Their algorithm and their analysis were based on the LP relaxation (L.P.1).
%
%
Seb\H{o} \cite{sebo13} presented a new analysis of this randomized algorithm
and improved the approximation factor to $\frac{8}{5}$.
The algorithm and analysis of \cite{sebo13} were based on
a different LP relaxation, see (L.P.4) in Section \ref{sec:LPs}.
In Section \ref{sec:LPs}, we prove that
(L.P.1) and (L.P.4) have the same optimal value.
This result together with a few more observations implies that
(L.P.4) can be replaced by (L.P.1) in
the algorithm and analysis of \cite{sebo13}
to achieve the same approximation factor of $\frac{8}{5}$.
In this section,
we prove the approximation factor of \cite{AKS12}; also,
we prove the $\frac{8}{5}$-approximation factor of \cite{sebo13}
based on (L.P.1) rather than (L.P.4).



\bigskip

\noindent \textbf{Randomized Christofides' algorithm:} \\
Solve the LP relaxation (L.P.1) to get an optimal solution $x^*$. Since $x^*$ is in the spanning tree polytope,
there exists a convex decomposition of spanning trees
$J_1, J_2, \ldots, J_l$ such that
$\sum_{1\leq i\leq l}\lambda_i \mathcal{X}^{J_i} = x^*$ where $\sum_{1\leq i\leq l}\lambda_i =1$, $\lambda_i>0$ and $\mathcal{X}^{J_i}$
is the edge incidence vector of $J_i$.
Such a decomposition can be found in polynomial time, see
Theorem 51.5 of \cite{Sch03}. We sample a spanning tree $J$ from these spanning trees according to
the probability defined by the coefficient $\lambda_i$ of
each spanning tree in the convex combination.
Let $T$ denote the set of the wrong degree vertices of $J$.
Then, as in the Christofides' algorithm,
a minimum-cost $T$-join $F$ is added to fix
the wrong degree vertices of $J$.

The expected cost of the random solution of the algorithm
is the sum of the expected cost of $J$, which is the cost of $x^*$,
and the expected cost of the $T$-join $F$.
Any feasible solution of the $T$-join polyhedron provides
a cost upper bound for the $T$-join $F$.
An et al. \cite{AKS12} introduced correction vectors to construct a special type of fractional $T$-join. A \emph{correction vector} for a $\tau$-narrow cut $S$ is an edge vector $z$ that satisfies $\sum_{e\in \delta(S)}z_e\geq 1$, where the definition of $\tau$-narrow cut will be given next. The correction vectors were further analyzed in \cite{sebo13}
to obtain a better approximation factor. In this section, we present a unified correction vector
to derive the results of both \cite{AKS12} and \cite{sebo13}.

The following key definition is introduced in \cite{AKS12}.
Let $0<\tau\leq 1$. If an $s$-$t$ cut $Q$ satisfies
$x^*(\delta(Q))<1+\tau$, we call it a \emph{$\tau$-narrow cut}.
Let $\mathcal{C}_\tau$ be the set of all $\tau$-narrow cuts
that contain $s$. It turns out that $\tau$-narrow cuts have
a nice structural property.

\begin{lemma}\cite{AKS12} \label{lem:nestedCuts}
Let $Q_1$, $Q_2$ be two distinct cuts in $\mathcal{C}_\tau$.
Then either $Q_1\subsetneq Q_2$ or $Q_2\subsetneq Q_1$.
\end{lemma}

For the sake of completeness, we present a proof.
\vspace{3mm}

\begin{proof}
Suppose that the statement is false.
Then both $Q_1\backslash Q_2$ and $Q_2\backslash Q_1$ are nonempty.
Note that both $Q_1\backslash Q_2$ and $Q_2\backslash Q_1$ are
$\{s,t\}$-even. Hence,
$x^*(\delta(Q_1))+x^*(\delta(Q_2))\geq
	x^*(\delta(Q_1\backslash Q_2))+x^*(\delta(Q_2\backslash Q_1))\geq 4$
by the constraints in (L.P.1).
However, $x^*(\delta(Q_1))+x^*(\delta(Q_2))< 2+2\tau \leq 4$.
This is a contradiction.
\end{proof}

Thus, we can use $Q_1, Q_2, \ldots, Q_k$ to denote all of
the $\tau$-narrow cuts containing $s$ such that
$s\in Q_1 \subsetneq Q_2 \subsetneq Q_3 \cdots \subsetneq Q_k\subsetneq V$.
Note that $\mathcal{C}_\tau=\{Q_i\}_{1\leq i\leq k}$.
Define $L_i=Q_i\backslash Q_{i-1}$ for $i=1, 2, \ldots, k, k+1$
where $Q_0=\emptyset$ and $Q_{k+1}=V$.
Each $L_i$ is nonempty and $\cup_{1\leq i\leq k+1}L_i=V$.
We call $\{L_i\}$ \emph{the partition derived by
the $\tau$-narrow cuts $\mathcal{C}_\tau$}.

Let $\mathcal{X}^J$ denote the edge incidence vector of the edge~set of $J$.
For any $Q\in \mathcal{C}_\tau$, we let $e_{Q}$ be
an edge in $\delta(Q)$ of minimum cost.
Let $\mathcal{X}^{e_{Q}}$ denote the edge incidence vector of $\{e_{Q}\}$,
i.e., $\mathcal{X}^{e_{Q}}_{e_{Q}}=1$, and $\mathcal{X}^{e_{Q}}_{e}=0$ if $e\neq e_Q$.
Our \emph{unified correction vector} is defined as $\mathcal{X}^{e_{Q}}$ for each $Q\in \mathcal{C}_\tau$,
i.e., the unified correction vector simply assigns
the value one to the minimum-cost edge in each $\tau$-narrow cut. In contrast, the correction vectors used
in \cite{AKS12} and \cite{sebo13} are fractional but sum up to at least one for each $\tau$-narrow cut.

Let $\alpha, \beta$ and $\tau$ be real parameters between $0$ and $1$,
whose specific values are given later.
Recall that $J$ is the random spanning tree in the
randomized Christofides' algorithm. Our fractional feasible $T$-join solution with unified correction
vectors, called \emph{unified fractional $T$-join}, is as follows:

\medskip

\noindent \textbf{Unified fractional $T$-join: }	
\[ f=\alpha \mathcal{X}^J + \beta x^* + \sum_{Q\in \mathcal{C}_\tau,\ Q\ \textnormal{is} \ T\textnormal{-odd}}
	(1-2\alpha - \beta x^*(\delta(Q)))\mathcal{X}^{e_{Q}}.
\]
where $\alpha, \beta, \tau$ satisfy the following condition:
\begin{equation}\label{settings}
\alpha+2\beta=1, \tau=\frac{1-2\alpha}{\beta}-1, \alpha\geq 0 \mbox{ and } \beta\geq 0.
\end{equation}

Let us derive the settings of $\alpha, \beta$ and $\tau$ in (\ref{settings}).
The purpose of the unified fractional $T$-join $f$ is to provide an upper bound on the cost of the minimum-cost $T$-join $F$ in the
randomized Christofides' algorithm. By Lemma \ref{lem:Tjoin}, it suffices to make $f$ feasible
for the $T$-join polyhedron (L.P.3). This requires special settings of $\alpha, \beta$ and $\tau$.

Consider the cut constraints in (L.P.3). Let $S$ be a $T$-odd cut. First we need to make sure that for any
$Q\in \mathcal{C}_\tau$, the coefficient $1-2\alpha - \beta x^*(\delta(Q))$ is nonnegative. Since $x^*(\delta(Q))<1+\tau$ for any $Q\in \mathcal{C}_\tau$, it
suffices to set $1-2\alpha -\beta(1+\tau)=0$, i.e., $\tau=\frac{1-2\alpha}{\beta}-1$.

Suppose that $S$ is an $s$-$t$ cut.
Note that $S$ is $T$-odd.
Hence, by Lemma \ref{lem:simpleParity}, $|\delta(S)\cap J|$ is even.
If $S$ is not a $\tau$-narrow cut, then
$f(\delta(S))\geq \alpha \mathcal{X}^J (\delta(S))+ \beta x^*(\delta(S)) \geq
2\alpha +\beta(1+\tau)$. By the assumption that $\tau=\frac{1-2\alpha}{\beta}-1$, we have $f(\delta(S))\geq 1$ in this case.
If $S$ is a cut in $\mathcal{C}_\tau$, then $f(\delta(S))\geq 2\alpha +\beta x^*(\delta(S)) +
(1-2\alpha - \beta x^*(\delta(S)))\mathcal{X}^{e_{S}}(\delta(S))\geq 1 $.

Now the only remaining case is that $S$ is $\{s, t\}$-even. Then $x^*(\delta(S))\geq 2$ by (L.P.1).
Since $J$ is a spanning tree, we have $\mathcal{X}^J(\delta(S))\geq 1$.
This implies $f(\delta(S))\geq
\alpha \mathcal{X}^J (\delta(S))+ \beta x^*(\delta(S)) \geq \alpha+ 2\beta$. Hence, in this case,
it suffices to set $\alpha+ 2\beta=1$.

Hence, we have the following result by the analysis above.

\begin{lemma}\label{lem:feasible}
The unified fractional $T$-join $f$ is
a feasible solution of the $T$-join polyhedron (L.P.3).
\end{lemma}

Lemma \ref{lem:feasible} shows that the expected cost of the minimum-cost $T$-join $F$ computed by
the randomized Christofides' algorithm is
at most the expected cost of the unified fractional $T$-join. Hence, the expected cost of the solution of the randomized
Christofides' algorithm is upper bounded by the optimal value of (L.P.1) plus the expected cost of the
unified fractional $T$-join.  In Section \ref{sec:AKS'results} and Section \ref{sec:sebo's results}, we will present two different analyses of the
expected cost of the unified fractional $T$-join to derive two different approximation factors from \cite{AKS12} and \cite{sebo13} for the randomized
Christofides' algorithm.

\begin{remark}
From the analysis above,  the cost analysis of the
unified fractional $T$-join is critical for proving
an approximation factor for the randomized Christofides' algorithm.
If we can get a better upper bound on the cost of the unified fractional $T$-join,
then the approximation factor can be further improved.
\end{remark}

The following lemma is used in the analysis of the expected cost of the unified fractional $T$-join in
Section \ref{sec:AKS'results} and Section \ref{sec:sebo's results}.

\begin{lemma}\cite{AKS12}\cite{sebo13}\label{lem:probBound}
Let $J$ be the random spanning tree and
$T$ be the set of wrong degree vertices of $J$ in the
randomized Christofides' algorithm.
Let $Q\in \mathcal{C}_\tau$, i.e., $Q$ is a $\tau$-narrow cut.
Then
\begin{itemize}
\item[(i)]
$\Pr(|\delta(Q) \cap J|=1) \geq 2-x^*(\delta(Q))$, and
\item[(ii)]
$\Pr(Q\ \textnormal{is} \ T\textnormal{-odd}) \leq x^*(\delta(Q))-1$.
\end{itemize}
\end{lemma}

\vspace{2mm}

For the sake of completeness, we present a proof.
\vspace{3mm}

\begin{proof}
Since $J$ is a spanning tree, $|\delta(Q) \cap J|\geq 1$ always holds.
So $\sum_{i\geq 1} \Pr(|\delta(Q) \cap J|=i)=1$. Then
\begin{eqnarray}
\Pr(|\delta(Q) \cap J|\geq 2)&\leq& \sum_{i\geq 1} i*\Pr(|\delta(Q) \cap J|= i)-\sum_{i\geq 1} \Pr(|\delta(Q) \cap J|=i)\nonumber \\
                             &=& \mathbb{E}(|\delta(Q) \cap J|)-\sum_{i\geq 1} \Pr(|\delta(Q) \cap J|=i)\nonumber \\
    &=&\quad x^*(\delta(Q))-1. \nonumber
\end{eqnarray}
Note that $\mathbb{E}(|\delta(Q) \cap J|)=x^*(\delta(Q))$ follows from the fact that $\mathbb{E}(\mathcal{X}^J)=x^*$ since $J$ is a random tree in the convex decomposition
of spanning trees for $x^*$ where the coefficients of the spanning trees define the probability distribution. Thus, we have
$\Pr(|\delta(Q) \cap J|=1) =
1-\Pr(|\delta(Q) \cap J|\geq 2)\geq 2-x^*(\delta(Q))$.
This proves the first inequality.

Now consider the second inequality.
By Lemma \ref{lem:simpleParity}, $|\delta(Q) \cap J|$ is even if $Q$ is $T$-odd.
This means $\Pr(Q\ \textnormal{is} \ T\textnormal{-odd}) \leq
\Pr(|\delta(Q) \cap J| \ \textnormal{is} \ \textnormal{even}) \leq
\Pr(|\delta(Q) \cap J|\geq 2) \leq x^*(\delta(Q))-1$.
\end{proof}


\subsection{AKS' $\frac{1+\sqrt{5}}{2}$-approximation via
unified correction vector} \label{sec:AKS'results}

First, we present two lemmas needed for the cost analysis of the randomized Christofides' algorithm.
\begin{lemma}\label{base_inj}
Let $K$ be a spanning tree with $n$ vertices. Let $\mathcal{S}=\{S_i:1\leq i\leq n-1\}$ be a family of subsets
of the vertex set of $K$ such that
$|S_i|=i$ and $S_i\subsetneq S_{i+1}$. There exists a bijection from $\mathcal{S}$ to $E(K)$ such that
each cut $S_i$ is mapped to an edge of $K$ in $\delta(S_i)$.
\end{lemma}

\begin{proof}Without loss of generality, we can assume that the vertex set of $K$ is $\{v_1, v_2, \ldots, v_n\}$
and $S_i=\{v_1, v_2, \ldots, v_i\}$ for $1\leq i\leq n-1$. We prove the result by induction on $n$.
The statement is clearly true for $n=2$.
Suppose $n\geq 3$.
Consider the vertex $v_n$.

We first pick the edge $e$ of $K$ incident with $v_n$
in the unique path of $K$ between $v_{n-1}$ and $v_n$. We map $S_{n-1}$ to this edge $e$. Let $K'$ be the graph obtained from $K\bs\{e\}$ by
contracting $v_{n-1}$ and $v_n$ into a single vertex $v'_{n-1}$. Note that $K'$ is a connected graph with $n-2$ edges. This implies
that $K'$ is a spanning tree with $n-1$ vertices $\{w_1, w_2, \ldots, w_{n-1}\}$ where $w_i=v_i$ for $1\leq i\leq n-2$ and $w_{n-1}=v'_{n-1}$.
Note that $\delta(\{w_1, w_2, \ldots, w_i\})$ is a subset of $\delta(S_i)$ for $1\leq i\leq n-2$. Hence, we can define
the rest of the bijection by applying the induction hypothesis
to the spanning tree $K'$ on these $n-1$ vertices.





\end{proof}

\begin{lemma}\label{lem:ineq1}
\begin{equation} \label{ineq1}
\sum_{Q\in \mathcal{C}_\tau} c(e_Q) \leq c(x^*)
\end{equation}
\end{lemma}

\begin{proof}
Let $K_{min}$ be a minimum-cost spanning tree on $G$.
Consider the partition $\{L_i\}$ derived by $\mathcal{C}_\tau$.
We contract every $L_i$ into a single vertex.
Then the resulting graph obtained from $K_{min}$ is connected.
Let $K$ be a spanning tree of the contracted graph.
Applying Lemma \ref{base_inj} to $K$,
we construct an injective mapping $\phi$ from $\mathcal{C}_\tau$
to the edge set of $K$ such that $\phi(Q)\in \delta(Q)$
for each $Q\in \mathcal{C}_\tau$.
Note that $K\subseteq K_{min}$.
Then $\sum_{Q\in \mathcal{C}_\tau} c(e_Q) \leq
\sum_{Q\in \mathcal{C}_\tau} c(\phi(Q)) \leq c(K_{min})\leq c(x^*)$
since $x^*$ is in the spanning tree polytope. The first inequality follows from the fact that
$e_Q$ is the minimum-cost edge in $\delta(Q)$.
\end{proof}

\begin{theorem}\cite{AKS12}\label{thm:AKS}
The randomized Christofides' algorithm achieves an
approximation factor of $\frac{1+\sqrt{5}}{2}$.
\end{theorem}
\begin{proof}
Since $J$ is a random spanning tree based on the convex decomposition of spanning trees for $x^*$, we have $\mathbb{E}(\mathcal{X}^J)=x^*$. Hence, the expected cost of the solution of the randomized
Christofides' algorithm is upper bounded by the optimal value of (L.P.1) plus the expected cost of the
minimum-cost $T$-join $F$. By Lemma \ref{lem:Tjoin} and Lemma \ref{lem:feasible}, the expected cost of $F$ is
at most the expected cost of the unified fractional $T$-join.
\begin{eqnarray}
&&\mathbb{E}[ c(\alpha \mathcal{X}^J+ \beta x^* + \sum_{Q\in \mathcal{C}_\tau,\ Q\ \textnormal{is} \ T\textnormal{-odd}} (1-2\alpha -\beta x^*(\delta(Q)))\mathcal{X}^{e_{Q}}) ] \nonumber \\
 &\stackrel{Lemma\ \ref{lem:probBound}}{\leq}& (\alpha+\beta)c(x^*)+\sum_{Q\in \mathcal{C}_\tau}(x^*(\delta(Q))-1)(1-2\alpha -\beta x^*(\delta(Q)))c(e_Q) \nonumber \\
 &\leq & (\alpha+\beta)c(x^*)+  \max_{0\leq z<\tau}z(1-2\alpha-\beta z-\beta)\sum_{Q\in \mathcal{C}_\tau}c(e_Q) \nonumber \\
  &\stackrel{Lemma \ \ref{lem:ineq1}}{\leq}&(\alpha+\beta+\max_{0\leq z<\tau}z(1-2\alpha-\beta z-\beta))c(x^*)\nonumber \\
  &\stackrel{ By \ (\ref{settings})}{=}&(\alpha+\beta+\beta\max_{0\leq z<\tau}z(\tau-z))c(x^*).\nonumber
\end{eqnarray}
The last equality follows from the fact that $1-2\alpha=\beta(\tau+1)$ by (\ref{settings}). The value of $z$ that maximizes
the expression is $\frac{\tau}{2}$. Hence, the upper bound on the expected cost of the unified fractional
$T$-join is at most $(\alpha + \beta + \beta(\frac{\tau}{2})^2)c(x^*)$. Substitute $\tau=\frac{1-2\alpha}{\beta}-1$, $\alpha =1- 2\beta$ from (\ref{settings}) into the upper bound.
Minimizing with respect to $\beta$ gives $\frac{\sqrt{5}-1}{2}c(x^*)$ with optimal settings:  $\beta=\frac{1}{\sqrt{5}}$,
$\alpha=1-\frac{2}{\sqrt{5}}$, $\tau=3-\sqrt{5}$.
Therefore, the optimal value of (L.P.1) plus this upper bound $\frac{\sqrt{5}-1}{2}c(x^*)$ on the expected cost of the unified fractional $T$-join leads to
the approximation factor of
$\frac{1+\sqrt{5}}{2}$ that was first proved in \cite{AKS12}.
\end{proof}

In \cite{AKS12}, the correction vector is constructed by using flow
computations to map the optimal LP solution $x^*$ to the $\tau$-narrow cuts. In contrast, our unified correction vector simply assigns the
value one to the minimum-cost edge in each $\tau$-narrow cut. We avoid the flow computation argument of \cite{AKS12} by using Lemma \ref{base_inj}.


\subsection{Seb\H{o}'s $\frac{8}{5}$-approximation via
unified correction vector}\label{sec:sebo's results}

Let $P$ be the $s$-$t$ path in $J$.
Seb\H{o} \cite{sebo13} points out the crucial fact that
$J\backslash P$ is a $T$-join for the set of wrong degree vertices $T$ of $J$. Recall that $F$ is the
minimum-cost $T$-join in the randomized Christofides' algorithm. This implies that $\mathbb{E}(c(F))\leq \mathbb{E}(c(J\backslash P))$.
Note that $c(x^*)=\mathbb{E}(c(J))=
\mathbb{E}(c(J\backslash P))+\mathbb{E}(c(P))$.

It turns out that $\mathbb{E}(c(P))$ also serves as
an upper bound in another cost inequality similar to (\ref{ineq1});
see the following lemma.

\begin{lemma}\label{lem:ineq2}
\begin{equation}\label{ineq2}
\sum_{Q\in \mathcal{C}_\tau} (2-x^*(\delta(Q)))c(e_Q) \leq \mathbb{E}(c(P)).
\end{equation}
\end{lemma}
\begin{proof}
Let $Q\in \mathcal{C}_\tau$; thus, $Q$ is a $\tau$-narrow cut.
If $|\delta(Q)\cap J|=1$,
then let $e'_Q$ denote the unique edge in $\delta(Q)\cap J$.
Recall that a $\tau$-narrow cut is an $s$-$t$ cut, and therefore
$e'_Q$ must be in $P$ since $P$ is the $s$-$t$ path in $J$.
Moreover, observe that $Q$ is one of the two connected components of
$J\bs\{e'_Q\}$.
Hence, for distinct $Q_1, Q_2 \in \mathcal{C}_\tau$
such that $|\delta(Q_1)\cap J|=1$ and $|\delta(Q_2)\cap J|=1$,
the edges $e'_{Q_1}$ and $e'_{Q_2}$ must be distinct
(otherwise, $J\bs\{e'_{Q_1}\}$ and $J\bs\{e'_{Q_2}\}$ would have
the same connected components, contradicting the fact that $Q_1,Q_2$
are distinct sets containing $s$).
Then
\[c(P)\geq \sum_{|\delta(Q)\cap J|=1, Q\in \mathcal{C}_\tau}c(e'_Q)\geq
\sum_{|\delta(Q)\cap J|=1, Q\in \mathcal{C}_\tau}c(e_Q).\]
By Lemma  \ref{lem:probBound},
\[\mathbb{E} (c(P)) \geq
\sum_{Q\in \mathcal{C}_\tau}\Pr( |\delta(Q)\cap J|=1) c(e_Q) \geq
\sum_{Q\in \mathcal{C}_\tau} (2-x^*(\delta(Q)))c(e_Q).\]
\end{proof}

\begin{theorem}\cite{sebo13}
The randomized Christofides' algorithm achieves an
approximation factor of $\frac{8}{5}$.
\end{theorem}
\begin{proof}
By an argument similar to the one in the proof of Theorem \ref{thm:AKS}, we are only concerned with the expected cost of the unified fractional $T$-join, which bounds the expected cost of the minimum-cost $T$-join $F$ in the randomized Christofides' algorithm.
\begin{eqnarray}\label{inq:rawuppbound}
&&\mathbb{E}[ c(\alpha \mathcal{X}^J+ \beta x^* + \sum_{Q\in \mathcal{C}_\tau,\ Q\ \textnormal{is} \ T\textnormal{-odd}} (1-2\alpha -\beta x^*(\delta(Q)))\mathcal{X}^{e_{Q}}) ] \nonumber \\
 &\stackrel{Lemma\ \ref{lem:probBound}}{\leq}& (\alpha+\beta)c(x^*)+\sum_{Q\in \mathcal{C}_\tau}(x^*(\delta(Q))-1)(1-2\alpha -\beta x^*(\delta(Q)))c(e_Q) \nonumber \\
 &\leq & (\alpha+\beta)c(x^*)+\sum_{Q\in \mathcal{C}_\tau}\frac{(x^*(\delta(Q))-1)(1-2\alpha -\beta x^*(\delta(Q)))}{2-x^*(\delta(Q))}(2-x^*(\delta(Q)))c(e_Q) \nonumber \\
 &\leq& (\alpha+\beta)c(x^*)+ \max_{0\leq z<\tau}\frac{z(1-2\alpha-\beta z-\beta)}{1-z} \sum_{Q\in \mathcal{C}_\tau}(2-x^*(\delta(Q)))c(e_Q) \nonumber \\
 &\stackrel{Lemma\ \ref{lem:ineq2}}{\leq}& (\alpha+\beta)c(x^*)+ \max_{0\leq z<\tau}\frac{z(1-2\alpha-\beta z-\beta)}{1-z}\mathbb{E}(c(P)) \nonumber \\
 &\stackrel{By\ (\ref{settings})}{=}& (\alpha+\beta)c(x^*)+ \beta\max_{0\leq z<\tau}\frac{z(\tau-z)}{1-z}\mathbb{E}(c(P)).
\end{eqnarray}
The last equality follows from the fact that $1-2\alpha=\beta(\tau+1)$ by (\ref{settings}). The value of $z$ that maximizes the expression is $1-\sqrt{1-\tau}$. Hence, the upper bound on the expected cost of the unified fractional
$T$-join is at most $(\alpha + \beta)c(x^*) + \beta(1-\sqrt{1-\tau})^2\mathbb{E}(c(P))$. Substitute $\tau=\frac{1-2\alpha}{\beta}-1$, $\alpha =1- 2\beta$ from (\ref{settings}) into (\ref{inq:rawuppbound}). Then the coefficients of the terms in (\ref{inq:rawuppbound}) only depend on $\beta$. Denote the coefficient of the last term in (\ref{inq:rawuppbound}) by $h(\beta)$ where $h(\beta)=(\sqrt{\beta}-\sqrt{1-2\beta})^2$. Then the bound can be written as $(1-\beta)c(x^*)+h(\beta)\mathbb{E}(c(P))$.
Note that $c(x^*)=\mathbb{E}(c(J\backslash P))+\mathbb{E}(c(P))$. Assume $\mathbb{E}(c(P))=\lambda_0 c(x^*)$. So $0\leq \lambda_0 \leq 1$
and $\mathbb{E}(c(J\backslash P))=(1-\lambda_0)c(x^*)$. Since $\mathbb{E}(c(J\backslash P))\geq \mathbb{E}(c(F))$, we have
\begin{eqnarray}\label{inq:maxmin}
\mathbb{E}(c(F))&\leq& \min \{(1-\lambda_0)c(x^*), (1-\beta+h(\beta)\lambda_0) c(x^*)\} \nonumber \\
                &\leq& \max_{0\leq \lambda \leq 1} \{\min \{(1-\lambda)c(x^*), (1-\beta+h(\beta)\lambda) c(x^*) \}\}.
\end{eqnarray}
$\lambda$ maximizes the expression when $(1-\lambda)c(x^*)=(1-\beta+h(\beta)\lambda)c(x^*)$. So $\lambda=\frac{\beta}{h(\beta)+1}$. Minimizing
the upper bound in (\ref{inq:maxmin}) with respect to $\beta$ gives $\frac{3}{5}c(x^*)$ with optimal settings:
 $\beta=\frac{4}{9}, \alpha=\frac{1}{9}, \tau=\frac{3}{4}$; moreover, $\lambda=\frac{2}{5}$.
Therefore, the optimal value of (L.P.1) plus this upper bound $\frac{3}{5}c(x^*)$ leads to
the approximation factor of
$\frac{8}{5}$ that was first proved in \cite{sebo13}.
\end{proof}


\section{Linear programming relaxations of the $s$-$t$ path TSP}\label{sec:LPs}

In this section, we investigate the relationship between two different LP
relaxations of the $s$-$t$ path TSP. Let $H=(V, E(H))$ be a connected
graph with nonnegative edge costs $c'$, and
let $s$ and $t$ be two fixed vertices.
For a partition $\mathcal{W}=\{W_1, W_2, \ldots, W_{\ell}\}$ of
the vertex set $V$,
let $\delta(\mathcal{W})$ denote $\cup_{1\leq i\leq \ell}\delta(W_i)$.
Let $G=(V, E)$ be the metric completion of $H$ with metric costs $c$.
As mentioned in Section \ref{sec:ucv}, (L.P.1) is a linear
programming relaxation of the $s$-$t$ path TSP on $G$.  Let $2H$
be the graph obtained from $H$ by doubling every edge of $H$. The
$s$-$t$ path TSP on $G$ is equivalent to
the problem of finding a minimum-cost trail in $2H$ from $s$ to $t$
visiting every vertex at least once (multiple visits are allowed for the vertices but not the edges).
Thus, the problem is to find a minimum-cost connected
spanning subgraph of $2H$ with $\{s,t\}$ as the odd-degree vertex set.
%
%
Hence, the following (L.P.4) is another LP relaxation of
the $s$-$t$ path TSP.

\medskip
$
\begin{array}{rrcll}
\hbox{({\bf L.P.4})}\quad  {\rm minimize}: & \sum_{e \in E(H)} c'_ex_e & & \\
{\rm subject~to}: & x(\delta(\mathcal{W})) & \geq & |\mathcal{W}|-1 & \forall \mbox{ partition } \mathcal{W} \mbox{ of } V \\
& x(\delta(S)) & \geq & 2 & \forall \emptyset \subsetneq S \subsetneq V, |S \cap \{s, t\}| \mbox{ even } \\
& x_e & \geq  & 0 & \forall e \in E(H)
\end{array}
$
Note that (L.P.4) is defined on the original graph $H$ but (L.P.1)
is defined on the metric completion $G$ of $H$.
%

\begin{figure}[h]
\begin{center}
  \includegraphics[scale=0.5]{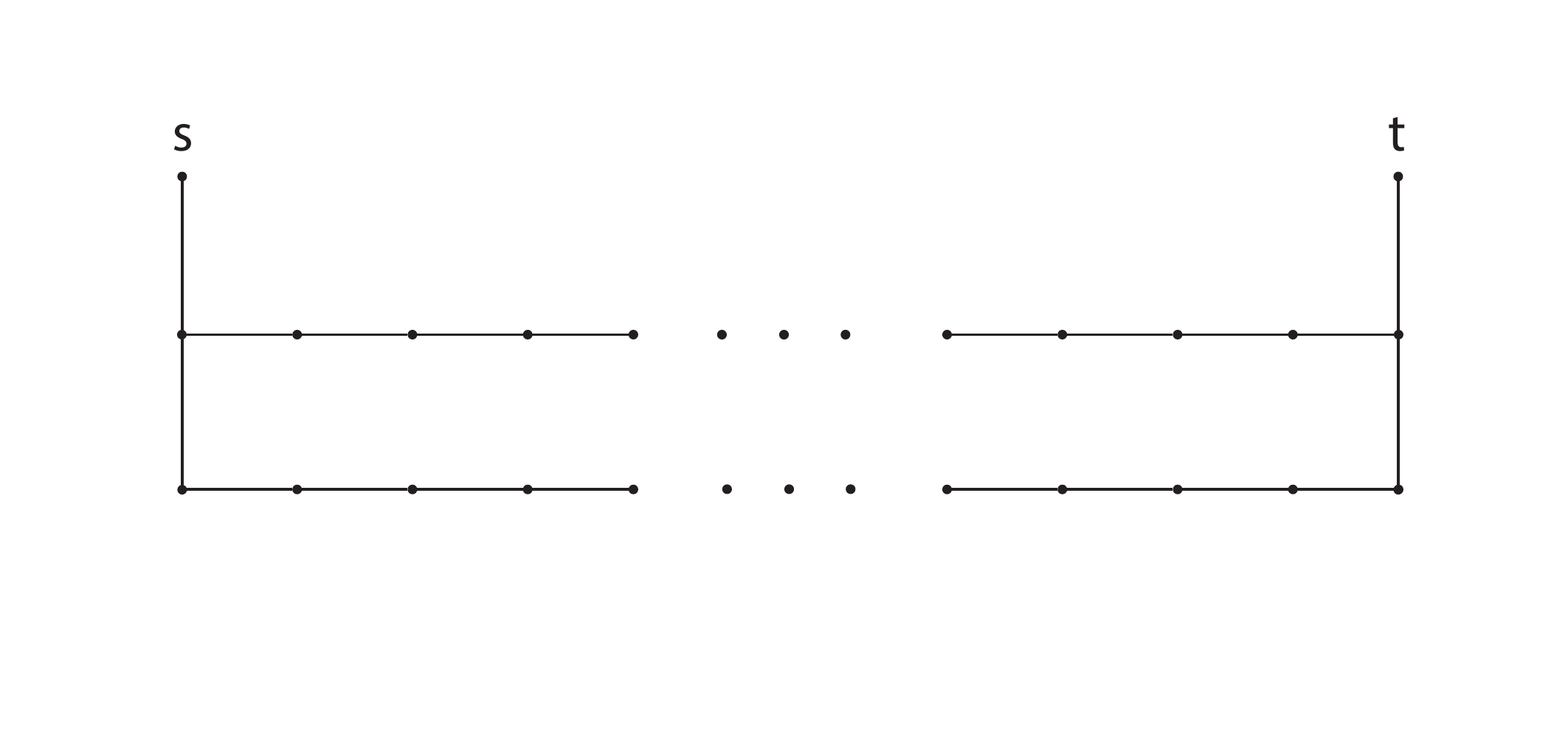}\\
  \caption{Tight Example}
  \label{tightExample}
\end{center}
\end{figure}

In this section, we show that both LPs, (L.P.1) and (L.P.4), have
the same (fractional) optimal value, see Corollary~\ref{cor:equiv}.
But these two LPs can differ with respect to integral solutions.
Observe that the integral solutions of (L.P.1) are
exactly the $s$-$t$ Hamiltonian paths of $G$;
this follows because an integral solution induces a graph
that is connected, has degree one at $s,t$, and has
degree two at all other vertices.
The integral solutions of (L.P.4) need not correspond to
the $s$-$t$ Eulerian paths of $H$; see the example shown
in Figure~\ref{tightExample}.

Let $Opt(LP_k)$ denote the optimal value  of (L.P.$k$), for $k=1,4$.
Let $Opt_{int}(LP_k)$ denote the minimum cost of
an integral solution that satisfies all constraints of (L.P.$k$),
for $k=1,4$.
We call $Opt_{int}(LP_k)$ the \emph{optimal integral value} of (L.P.$k$).
The following table summarizes the relationship between the two LPs;
the new results of this section appear in the last two columns.

\begin{center}
\setlength{\tabcolsep}{1pt}
\scalebox{0.75}{
\begin{tabular}{ | c | c | c | c | c |}
\hline
   LPs &  Graph & Costs & Optimum & Optimal Integral Value \\
\hline
  (L.P.1) & \small {$G$: metric completion of $H$ } & \small {$c$: metric extension of $c'$ }& $Opt(LP_1)$ & $Opt_{int}(LP_1)$ \\ \hline
  (L.P.4) & $H$ & $c'\geq 0$ &  $Opt(LP_1)$ & \small {$Opt_{int}(LP_4) \leq Opt_{int}(LP_1) \leq \frac{3}{2} Opt_{int}(LP_4)$} \\
  \hline
\end{tabular}
}
\end{center}

To obtain these results, we need an edge-splitting lemma.
Let $K$ be a multigraph, i.e., two adjacent vertices in $K$ may be
connected by one or more edges.
Let $(u, v), (v, w)\in E(K)$. The \emph{splitting operation} on
$(u, v), (v, w)$ at the vertex $v$ is defined as follows:

\begin{itemize}
\item Remove $(u, v), (v, w)$ and then add $(u, w)$ if $u\neq w$.
\end{itemize}

If $u=w$, then we remove the loop formed by adding $(u, w)$; note that
this removal of the loop has no effect on the edge-connectivity of the graph. We use the following
result to prove Lemma \ref{lem:17trans}; see \cite[Theorem $A'$]{Frank92}.

\begin{lemma}\cite{Lov74}\cite[Ex. 6.51]{Lov79}\label{lem:splitting}
Let $K$ be a multigraph with even degree at each vertex.
Let $v\in V(K)$ and let $U=V(K)\backslash \{v\}$.
Let $d$ be a positive integer. If

\begin{equation}\label{cond}
|\delta(S)|\geq d \text{ for each } \emptyset \subsetneq S \subsetneq U
\end{equation}
then the edges incident with $v$ can be partitioned into
$\frac{|\delta(v)|}{2}$ disjoint edge pairs $(p, v)$, $(v, q)$
such that the multigraph obtained by applying
the splitting operation to any one of these edge pairs (at the vertex $v$)
still satisfies (\ref{cond}).
\end{lemma}

\begin{lemma}\label{lem:17trans}
Let $x$ be a rational solution of (L.P.4) of cost $c'(x)$.
Then there exists a solution $x'$ of (L.P.1) with cost at most $c'(x)$.
Moreover, if $x$ is an integral solution, then $x'$ is half-integral.
\end{lemma}
\begin{proof}
The first part of this statement follows from the parsimonious property shown in \cite{BT97}. However, to show the second part of the statement, we present a proof for the first part as well.

Define an edge vector $y$ on $G$ as follows:
\begin{equation}
y_e = \begin{cases}
  x_e, & \text{if } e\in E(H), \\ \nonumber
  0, & \text{otherwise}.
\end{cases}
\end{equation}
Since $G$ is the metric completion of $H$, we know $c(y)\leq c'(x)$.
Then we construct $y'$ from $y$ as follows:

\begin{equation}
y'_e = \begin{cases}
  1+y_e, & \text{if } e=(s,t), \\ \nonumber
  y_e, & \text{otherwise}.
\end{cases}
\end{equation}

By the constraints of (L.P.4) and
the fact that $y'_{(s,t)}=y_{(s,t)}+1$,
we have $y'(\delta(S))\geq 2$ for each cut $S$.
Let $C$ be a positive integer such that $Cy'$ is integral.
Consider the multigraph $K_{2C}$ with $2Cy'_{(u, v)}$
number of edges between $u$ and $v$.
Then $|\delta_{K_{2C}}(S)|\geq 4C$.

By using Lemma \ref{lem:splitting}, we apply splitting operations at every
vertex until the degree of every vertex is exactly $4C$. We claim
that this procedure can be applied such that the number of edges
between $s$ and $t$ is $\geq 2C$. To see this, consider a splitting
operation at $s$ or $t$, say $s$; note that splitting at other vertices
does not decrease the number of edges between $s$ and $t$. There are at
least $2C+1$ feasible splitting pairs available at $s$ (since otherwise there is no
need to do a splitting operation at $s$, i.e., $|\delta_{K_{2C}}(s)|= 4C$). This implies
that we can always choose a splitting pair such that at least $2C$
edges between $s$ and $t$ are preserved.

Let $z$ be the edge vector associated with the resulting graph after
splitting, i.e., $z_{(u, v)}$ equals the number of edges between
$u$ and $v$ in the resulting graph. Furthermore, let $z'=z/2C$.
Then $z'(\delta(S))\geq 2$ for each cut $S$, $z'(\delta(v))=2$ for
each vertex $v$, and $z'_{(s, t)}\geq 1$. Consider two different
vertices $u, v$. We know $z'(\delta(u))=z'(\delta(v))=2$ and
$z'(\delta(\{u, v\}))\geq 2$. This implies $z'_{(u,v)}\leq 1$. In
particular, $z'_{(s, t)}=1$. Construct $x'$ from $z'$ as follows:

\begin{equation}
x'_e = \begin{cases}
  z'_e-1=0, & \text{if } e=(s,t), \\ \nonumber
  z'_e, & \text{otherwise}.
\end{cases}
\end{equation}

By the properties obtained for $z'$, we have $x'$ is a feasible
solution of (L.P.1). Note that the splitting operations never
increase the total cost since the edge costs are metric on $G$.
Therefore, the cost of $x'$ is at most $c'(x)$. In particular,
if $x$ is integral, we can set $C=1$ in the procedure.
In this case, $x'$ is half-integral.
\end{proof}

Conversely, any feasible solution of (L.P.1) can be transformed to
a feasible solution of $(L.P.4)$:
the idea is to replace each edge $(u,v)$ in $E(G)$
by a shortest $u$-$v$ path in $H$.
Note that every solution of (L.P.1) is a feasible solution of
the spanning tree polytope. Hence,
it can be seen that the transformed solution is feasible for (L.P.4),
and, in particular, it satisfies the partition constraints in (L.P.4).
Hence,
 \begin{equation}\label{LP4lesLP1}
 Opt(LP_4)\leq Opt(LP_1), \quad Opt_{int}(LP_4)\leq Opt_{int}(LP_1).
 \end{equation}

By Lemma \ref{lem:17trans}, we have the following result.

\begin{corollary}\label{cor:equiv}
$Opt(LP_4)=Opt(LP_1)$.
\end{corollary}

However, (L.P.1) and (L.P.4) may differ
in terms of the integral optimal value.
Consider the graph with unit edge costs in Figure~\ref{tightExample};
this is meant to be the original graph $H$ in the instance
of the $s$-$t$ path TSP.

Note that (L.P.4) is defined on the original graph but (L.P.1) is
defined on the metric completion. Let $\ell$ be the length of the
middle path in Figure \ref{tightExample}. It is not hard to see
that $Opt_{int}(LP_1)\approx 3\ell$ but $Opt_{int}(LP_4)\approx
2\ell$ when $\ell$ is sufficiently large. (For (L.P.4), consider the integral solution with value $1$ for
every edge of the original graph.) In this case,
$\frac{Opt_{int}(LP_1)}{Opt_{int}(LP_4)}\approx \frac{3}{2}$.
Interestingly, $\frac{3}{2}$ can be proved to be an upper bound for this
ratio. This example shows that the upper bound of $\frac{3}{2}$ is tight. To prove this upper bound, we present an algorithm to round a half-integral solution of (L.P.1) to an integral one by increasing the cost by a factor of at most $\frac{3}{2}$.

Apply the randomized Christofides' algorithm to a half-integral
solution $x$ of (L.P.1). Let $J$ be the random spanning tree obtained
from $x$. Let $F$ be a minimum-cost $T$-join for the set of wrong degree
vertices $T$ of $J$.

\begin{lemma}\label{lem:fea-half}
$x(\delta(S))\geq 2$ for any  $T$-odd cut $S$.
\end{lemma}
\begin{proof}
For any vertex $v\in V$, $x(\delta(v))$ is integral by
the constraints of (L.P.1).
Since $x_e$ is half-integral, $x(\delta(S))=\sum_{v\in S}
x(\delta(v))-2x(E(S))$ implies that $x(\delta(S))$ is integral.
Suppose $x(\delta(S))<2$ for some $T$-odd cut $S$. Then we have
$x(\delta(S))=1$. By the constraints of (L.P.1), $S$ must be an
$s$-$t$ cut. Note that $\mathbb{E}(\mathcal{X}^J)=x$ and $|J\cap \delta(S)|
\geq 1$ since $J$ is a random spanning tree. This implies $|J\cap
\delta(S)|=1$ always holds. However, since $S$ is an $s$-$t$ cut and
also a $T$-odd cut, we have $|\delta(S)\cap J|$ is even by Lemma \ref{lem:simpleParity}. This is a contradiction.
\end{proof}

\begin{theorem}\label{halfAlgorithm}
If the input is a half-integral solution $x$ of (L.P.1), then the
randomized Christofides' algorithm outputs a Hamiltonian $s$-$t$
path with cost at most $\frac{3}{2}c(x)$.
\end{theorem}
\begin{proof}
By Lemma \ref{lem:fea-half}, $\frac{1}{2}x$ is a feasible solution
of the $T$-join polyhedron (L.P.3).
This means $\mathbb{E}(c(F))\leq \frac{1}{2}c(x)$.
Therefore $\mathbb{E}(c(J))+\mathbb{E}(c(F))\leq \frac{3}{2}c(x)$.
\end{proof}

\medskip

Now we are ready to prove the ratio for the optimal integral values
of the two LPs.

\begin{theorem}\label{equivLP41}
$Opt_{int}(LP_4)\leq Opt_{int}(LP_1)\leq \frac{3}{2}Opt_{int}(LP_4)$.
Moreover, the bounds are tight.
\end{theorem}
\begin{proof}
The lower bound is due to (\ref{LP4lesLP1}). Now consider the upper
bound.  Let $x$ be an optimal integral solution of (L.P.4). By
Lemma \ref{lem:17trans}, there exists a half-integral solution $x'$
of (L.P.1) such that $c(x')\leq c'(x)$. By Theorem \ref{halfAlgorithm},
we can get an $s$-$t$ Hamiltonian path with cost at most $\frac{3}{2}c(x')$.
This means $Opt_{int}(LP_1)\leq \frac{3}{2}c(x')\leq
\frac{3}{2}c'(x)=\frac{3}{2}Opt_{int}(LP_4)$.

The tight example for the upper bound is shown in Figure
\ref{tightExample}. For the tightness of the lower bound, consider
the graph $H$ consisting of one path connecting $s$ and $t$ where
every edge has unit cost.
\end{proof}


\section{Counterexample to two approaches}\label{sec:BI}
For $s$-$t$ path TSP, the main question is whether there exists a $\frac{3}{2}$-approximation algorithm. When addressing this problem, two natural questions arise:

\begin{itemize}
\item \cite{Gao13} presented a simple $\frac{3}{2}$-approximation algorithm for the $s$-$t$ path TSP in the graphic case. Does it extend to give the same approximation factor for the general metric case?
\item Does every spanning tree in a given convex decomposition of an optimal solution $x$ of (L.P.1) achieve a $\frac{3}{2}$-approximation factor by adding a minimum-cost $T$-join to fix the wrong degree vertices ?
\end{itemize}

The first question concerns the extension of the algorithm for the graphic case. The second question focuses on the role of randomness and probabilistic methods in the analysis of the recent LP-based approximation algorithms. We answer these questions negatively by providing a counterexample.
In the following, we make the questions more precise and then show how our counterexample serves as a negative answer.

Let $G$ be the metric completion of some connected graph $H$ with unit edge costs $c'$.
The $s$-$t$ path TSP defined on $G$ is called $s$-$t$ path graph-TSP. In this important special case,
the gap between the upper bound and lower bound of the LP integrality ratio has been closed. The first $\frac{3}{2}$-approximation algorithm
for the $s$-$t$ path graph-TSP was given by \cite{SV12} using sophisticated techniques. \cite{Gao13} presented another $\frac{3}{2}$-approximation
algorithm which was conceptually simpler than that in \cite{SV12}.

Let $x^*$ be an optimal solution of the (L.P.4) defined on $H$. Note that $c'_e=1$ for $e\in E(H)$ in this case.
Let $Q$ be an $s$-$t$ cut. If $x^*(\delta(Q))<2$, we call it a \emph{narrow cut}, which is exactly a $1$-narrow
cut as defined in Section \ref{sec:ucv}. Note that the narrow cuts containing $s$ still have the nice structural
property of Lemma \ref{lem:nestedCuts} even when $x^*$ is an optimal solution of (L.P.4). We recall some notation
from Section \ref{sec:ucv}. The cuts $Q_1, Q_2, \ldots, Q_k$ are all the narrow cuts containing $s$ such that
$s\in Q_1 \subsetneq Q_2 \subsetneq Q_3 \cdots \subsetneq Q_k\subsetneq V$. Define $L_i=Q_i\backslash Q_{i-1}$
for $i=1, 2, \ldots, k, k+1$ where $Q_0=\emptyset$ and $Q_{k+1}=V$. Note that each $L_i$ is nonempty and $\cup_{1\leq i\leq k+1}L_i=V$.
It is shown in \cite{Gao13} that $H$ restricted on each $L_i$ is connected and also there exists at least one edge between each two
consecutive $L_i$ and $L_{i+1}$ in $H$.

We sketch the $\frac{3}{2}$-approximation algorithm in \cite{Gao13}. The algorithm constructs a minimal spanning tree on each $L_i$ and then connects them together by a unit cost edge between each two consecutive $L_i$ and $L_{i+1}$. This results in a spanning tree on $H$, which is called a \emph{good spanning tree}. Then a minimum-cost $T$-join $F_{good}$ is added to correct the wrong degree vertices of the good spanning tree. Since every edge in $H$ has unit cost, the good spanning tree has minimum cost, which is at most $Opt(LP_4)$. Furthermore, it is shown in \cite{Gao13} that the minimum-cost $T$-join $F_{good}$ has cost at most $\frac{1}{2}Opt(LP_4)$. This gives a $\frac{3}{2}$-approximation factor in total.

The only part in the analysis using the graphic property is
that the good spanning tree has cost at most $Opt(LP_4)$. A natural extension of the definition of a good spanning tree would be as follows:
\begin{itemize}
\item In the general metric case, a good spanning tree is constructed by connecting the minimum-cost spanning tree in each $L_i$ with a minimum-cost edge from $L_{i}$ to $L_{i+1}$.
\end{itemize}
If the cost of this ``extended" good spanning tree is bounded above by $Opt(LP_4)$ in the general metric case,
then it gives us a $\frac{3}{2}$-approximation factor for $s$-$t$ path TSP.
Unfortunately, this is not true. To show this, we present our counterexample, a complete graph $G=H=H_b$
with metric edge costs $c^{H_b}$ and vertex set $\{0, 1, \ldots, 7\}$ where $s=0, t=7$. The metric edge costs $c^{H_b}$ are given by the metric completion of the costs indicated in Figure \ref{optSoln} below. Note that for every edge $e$ in Figure \ref{optSoln}, $c^{H_b}_e$ is exactly the edge cost value shown in that figure.

Figure \ref{optSoln} shows the support graph of a feasible solution $x^{H_b}$ of (L.P.4),
where the first number on each edge denotes the $x^{H_b}$ value and the second number denotes the cost of the edge.

\begin{figure}[h]
\begin{center}
  \includegraphics[scale=0.5]{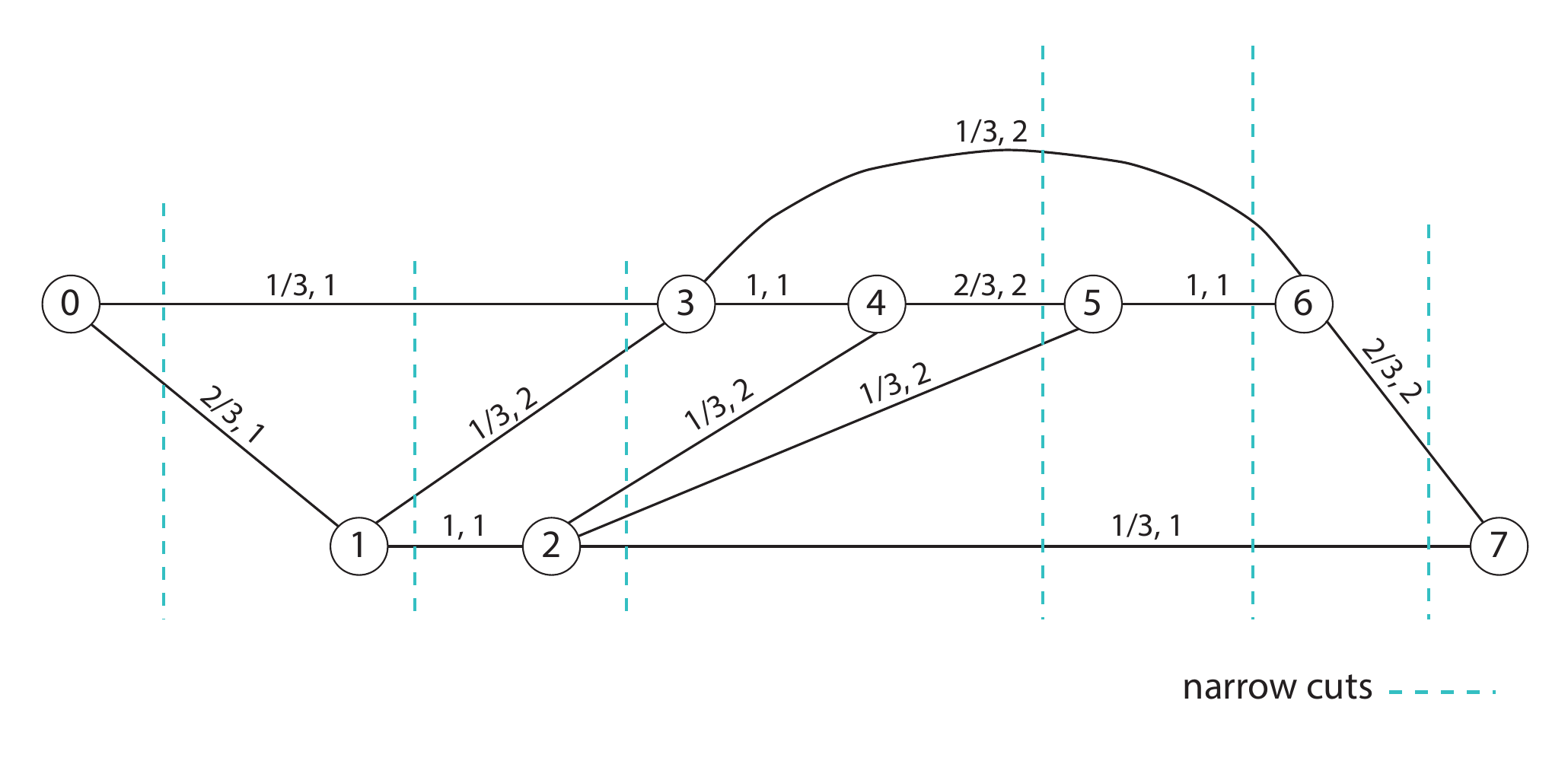}\\
  \caption{Support graph of $x^{H_b}$ with edge $x^{H_b}$ values and edge costs}
  \label{optSoln}
\end{center}
\end{figure}

\begin{lemma}\label{lem:feaExtreme}
$x^{H_b}$ is an optimal solution for (L.P.4) with respect to $c^{H_b}$.
Furthermore, $x^{H_b}$ is an extreme point of the polyhedron of (L.P.4) on $H_b$.
\end{lemma}
\begin{proof}
To show the optimality of $x^{H_b}$ for (L.P.4), it is sufficient to prove that $x^{H_b}$ is an optimal
solution of (L.P.1) by Corollary \ref{cor:equiv}. We use complementary slackness conditions to prove the optimality of $x^{H_b}$ for (L.P.1).
Let $\mathcal{S}_1$ be the set of all $s$-$t$ cuts and $\mathcal{S}_2$ be the set of all $\{s,t\}$-even cuts. Let $\mathcal{S}=\mathcal{S}_1\cup \mathcal{S}_2$.

$
     \begin{aligned}
     \hbox{({\bf Dual of (L.P.1)})}  \\
     {\rm maximize}: &\ y_s+y_t+2\sum_{v\notin\{s, t\}}y_v +\sum_{S\in \mathcal{S}_1}d_{S}+2\sum_{S\in \mathcal{S}_2}d_{S}-\sum_{e}u_e & \\
{\rm subject~to}:  \\
&y_w+y_v-u_{(w,v)}+\sum_{(w, v)\in \delta(S), S\in \mathcal{S}} d_{S} \leq c_{(w, v)},\ \ (w, v)\in E& \\
&u, d \geq 0  &
     \end{aligned}
$

The following dual solution $y, d, u$ witnesses the optimality of $x^{H_b}$ to (L.P.1) by the complementary slackness conditions:
\begin{itemize}
\item $u_{(1, 2)}=u_{(3, 4)}=\frac{2}{3}$, $u_{(5, 6)}=\frac{4}{3}$, and $u_{e}=0$ for any other edge $e$
\item $d_{\{3, 4, 5, 6\}}=\frac{1}{3}$ and $d_{S}=0$ for any other $S$
\item $y_{0}=0, y_2=y_3=\frac{2}{3}, y_1=y_4=y_5=1, y_6=\frac{4}{3}, y_7=\frac{1}{3}$
\end{itemize}
Hence $x^{H_b}$ is also an optimal solution of (L.P.4).

Denote the polyhedron of (L.P.4) on $H_b$ by $K$. We now show that $x^{H_b}$ is an extreme point of $K$. Otherwise,
there exists $x^{H_b}\neq z\in K$ and $z'\in K$ such that $x^{H_b}=\lambda z +(1-\lambda)z'$ for some
$0<\lambda <1$.

Clearly, for any edge $e$ not in the support graph of $x^{H_b}$, we have $z_e=0$ by Lemma \ref{lem:decomposition}.
We also
apply Lemma \ref{lem:decomposition} to $\delta(v)$ for each vertex $v$, and the cuts $S_1=\{3, 4\}, S_2=\{1, 2\}, S_3=\{5, 6\}, S_4=\{3, 4, 5, 6\}$.
Then, $z(\delta(v))=1$ for $v=0,7$ and $z(\delta(v))=2$ for other vertices, and $z(\delta(S_j))=2$ for $1\leq j \leq  4$.
Hence, $z_e=1$ for each $e\in E_1=\{(3, 4), (1, 2), (5, 6)\}$. Let $a=z_{(0, 3)}, b=z_{(4, 5)}$. By the $z$-values on the edges in $E_1$ and the values $z(\delta(v))$ for $v\in V(H_b)$, we have $z_{(0, 1)}=1-a, z_{(1, 3)}=a,  z_{(3, 6)}=1-2a, z_{(6, 7)}=2a, z_{(2,7)}=1-2a, z_{(2, 5)}=1-b, z_{(2, 4)}=1-b$.
Now consider $\delta(2)$ and $\delta(S_4)$. Then
\[ 2(1-b)+(1-2a)+1 =2, \quad 4a+2(1-b)=2.\]
Hence, $a=\frac{1}{3}, b=\frac{2}{3}$. By checking each edge, $z=x^{H_b}$. This is a contradiction. Therefore, $x^{H_b}$ is an extreme point of $K$. Note that
the analysis above also shows that $x^{H_b}$ is an extreme point of the polytope of (L.P.1) on $H_b$.

\end{proof}

\begin{figure}[h]
\begin{center}
  \includegraphics[scale=0.5]{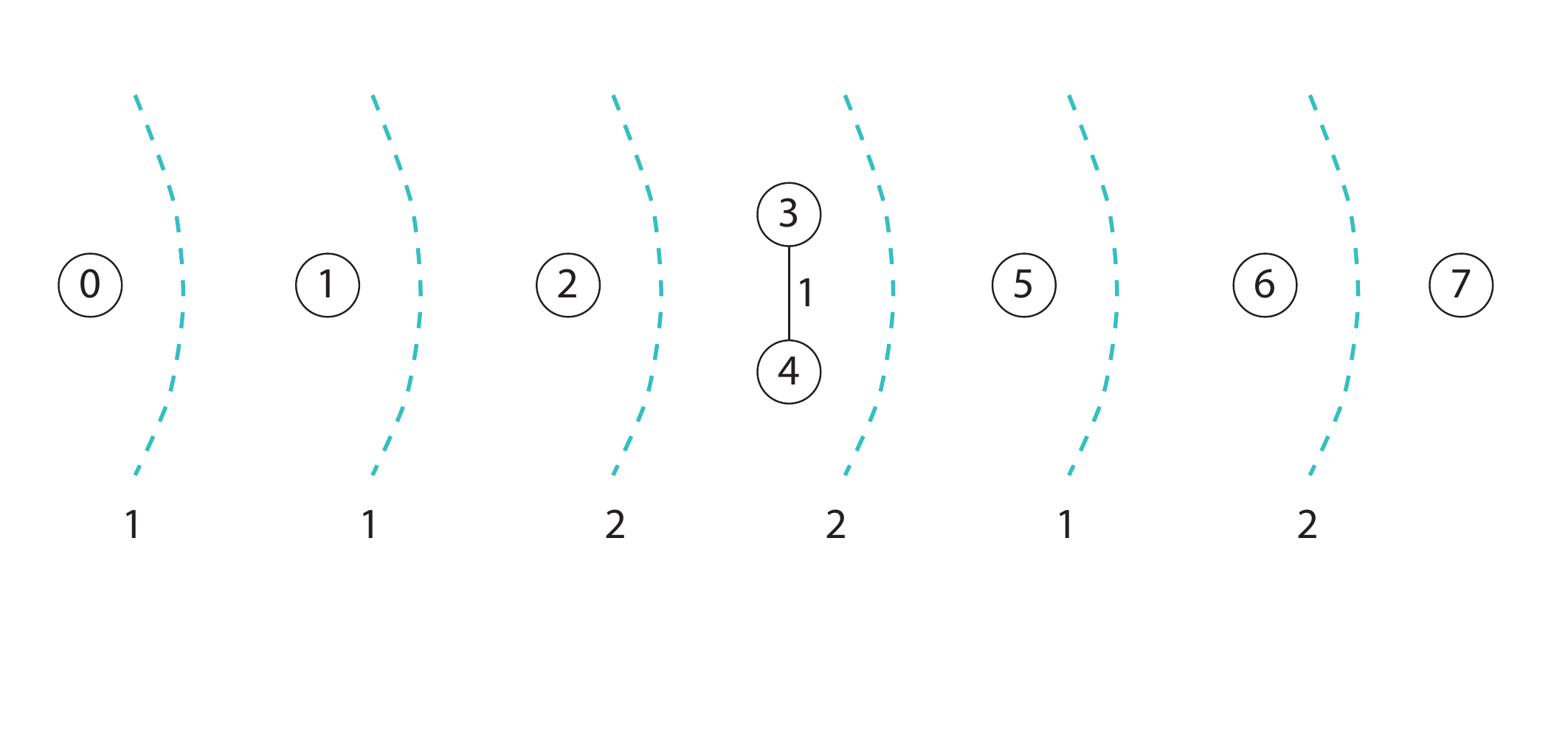}\\
  \caption{Cost of the good spanning tree}
  \label{COGST}
\end{center}
\end{figure}

The cost of the corresponding good spanning tree is $10$ and is shown in Figure \ref{COGST}.
The number on the edge between $3$ and $4$ in Figure \ref{COGST} is the edge cost.
The numbers below the dashed narrow cuts are the minimum costs of the edges crossing the
 narrow cuts to connect two consecutive parts. By Lemma \ref{lem:feaExtreme}, we know the optimal value of (L.P.4) is $c^{H_b}(x^{H_b})=9\frac{2}{3}$.
 So, we can see that the cost of the good spanning tree
 is strictly larger than the optimal value of (L.P.4). This refutes the statement that the cost of the ``extended"
good spanning tree can be upper bounded by $Opt(LP_4)$.

Interestingly, this instance also illustrates that probabilistic methods are
important for the analyses of improved LP-based approximation
algorithms such as the  ``randomized Christofides' algorithm" or its
deterministic version the ``best-of-many Christofides' algorithm"
(see \cite{AKS12}). The randomized Christofides' algorithm obtains a better approximation factor
by sampling a spanning tree $J$ from the convex decomposition of $x^*$.  However, is it true that for an arbitrary spanning tree in the support of a given convex decomposition,
the cost of the spanning tree plus a minimum-cost  $T$-join is at most $\frac{3}{2}Opt(LP_1)$ ?
In the rest of this section, via the instance $H_{b}$, we show this statement is false in general.

\begin{figure}[h]
\begin{center}
  \includegraphics[scale=0.5]{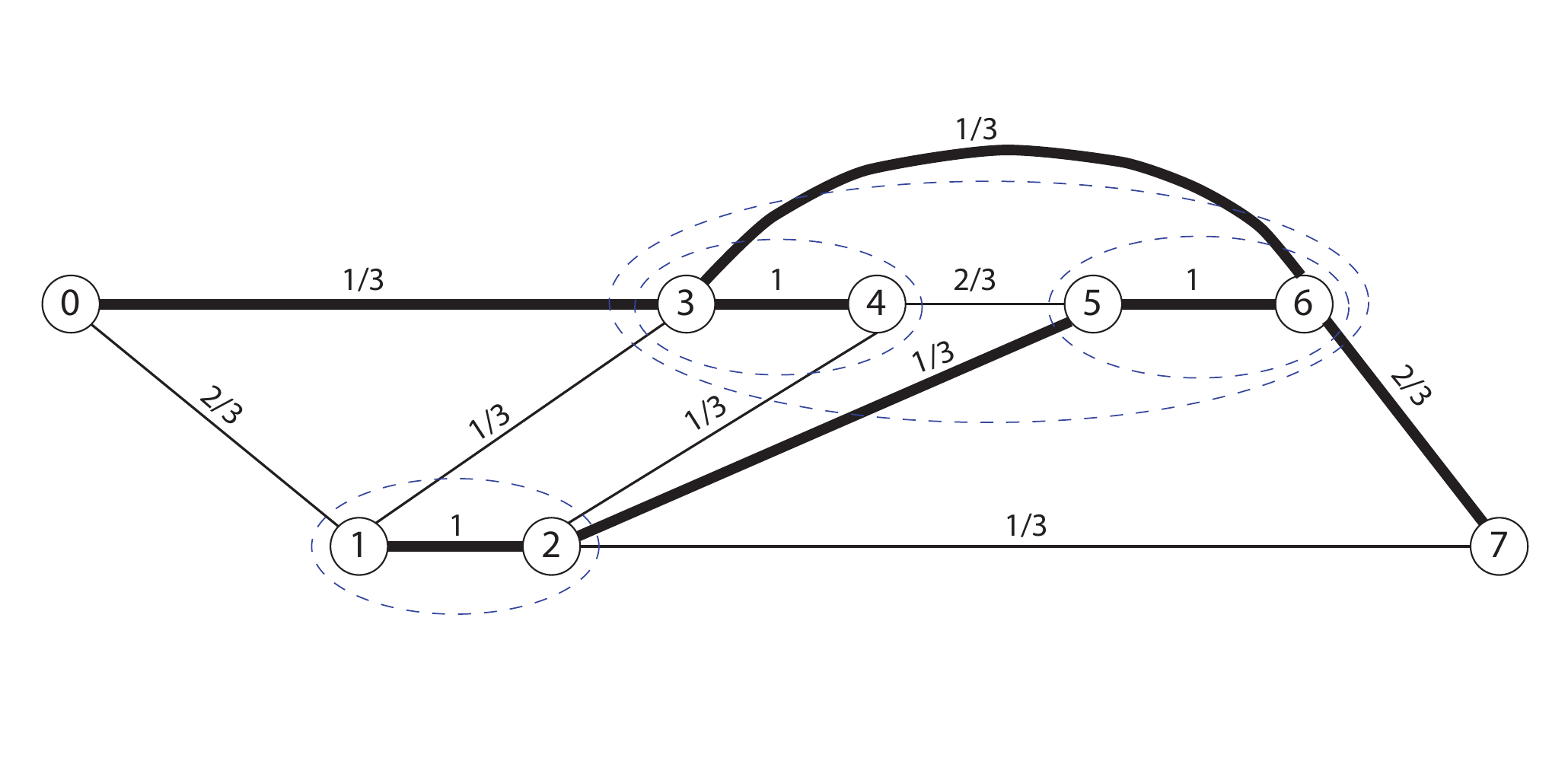}\\
  \caption{Tree $J_b$}
  \label{TreeJB}
\end{center}
\end{figure}

We recall the optimal solution $x^{H_b}$ of (L.P.1) on $H_b$ with metric costs $c^{H_b}$. We know that $x^{H_b}$ is in the spanning
tree polytope (L.P.2). The tight constraints of $x^{H_b}$ for the inequality constraints of (L.P.2) are illustrated as dashed circles in the Figure \ref{TreeJB} except
the tight constraints for $V\backslash \{s\}$, $V\backslash \{t\}$, $V\backslash \{s, t\}$.

By Lemma \ref{lem:decomposition}, the tree $J_b$ with the dark edges in the graph of Figure \ref{TreeJB} is in some convex decomposition of $x^{H_b}$ in (L.P.2),
i.e., $J_b$ is a spanning tree in the support of some convex decomposition of $x^{H_b}$. Let $T_b$ be the set of wrong degree vertices of $J_b$, i.e., $T_b=\{1, 3, 4, 6\}$. $F_b=\{(3, 6), (1, 4)\}$ is a minimum-cost $T_b$-join with cost $5$. Hence, the total cost
of the disjoint union of $J_b$ and $F_b$ is $15$, which is larger than $\frac{3}{2}$ times the optimal value $c^{H_b}(x^{H_b})=9\frac{2}{3}$ of (L.P.1). This shows the importance of the probabilistic techniques in the analysis of
the ``randomized Christofides' algorithm" or its
deterministic version the ``best-of-many Christofides' algorithm". Note that the minimum-cost $T_b$-join $F_b$ to
fix the wrong degree vertices of $J_b$ is also larger than half of the optimal value $9\frac{2}{3}$ of (L.P.1).

\medskip
\noindent
{\bf Acknowledgements}.
The author is grateful to Joseph Cheriyan for indispensable help, and to Zachary Friggstad for stimulating discussions.  Many thanks to Nishad Kothari, Andr\'{e} Linhares, Abbas Mehrabian, Andr\'{a}s Seb\H{o}, Chaitanya Swamy and Jens Vygen for their useful comments. The author also would like to thank the anonymous reviewers for their valuable advice for improving the presentation.


\bibliographystyle{alpha}
\bibliography{notes}

\begin{thebibliography}{CFG12}

\bibitem[AKS12]{AKS12}
H-C. An, R.~Kleinberg, and D.~B. Shmoys.
\newblock Improving {C}hristofides' algorithm for the s-t path {TSP}.
\newblock In {\em Proceedings of the 44th ACM Symposium on Theory of
  Computing}, pages 875--886, 2012.

\bibitem[BB08]{BB08}
G.~Benoit and S.~Boyd.
\newblock Finding the exact integrality gap for small {T}raveling {S}alesman
  {P}roblems.
\newblock {\em Mathematics of Operations Research}, 33(4):921--931, 2008.

\bibitem[BT97]{BT97}
D.~Bertsimas and C-P. Teo.
\newblock The parsimonious property of cut covering problems and its
  applications.
\newblock {\em Operations Research Letters}, 21(3):123--132, 1997.

\bibitem[CFG12]{CFG12}
Joseph Cheriyan, Zachary Friggstad, and Zhihan Gao.
\newblock Approximating minimum-cost connected {T}-joins.
\newblock In {\em Proceedings of the 15th International Workshop on
  Approximation Algorithms for Combinatorial Optimization Problems}, pages
  110--121, 2012.

\bibitem[Chr76]{christofides76}
N.~Christofides.
\newblock Worst-case analysis of a new heuristic for the {T}ravelling
  {S}alesman {P}roblem.
\newblock Technical report, Graduate School of Industrial Administration, CMU,
  1976.

\bibitem[EJ01]{EJ01}
J.~Edmonds and E.~L. Johnson.
\newblock Matching: A well-solved class of integer linear programs.
\newblock In {\em Combinatorial Optimization}, pages 27--30, 2001.

\bibitem[Fra92]{Frank92}
A.~Frank.
\newblock {O}n a theorem of {M}ader.
\newblock {\em Discrete Mathematics}, 101(1-3):49--57, 1992.

\bibitem[Gao13]{Gao13}
Z.~Gao.
\newblock An {LP}-based 3/2-approximation algorithm for the {\it s}-{\it t}
  path graph {T}raveling {S}alesman {P}roblem.
\newblock {\em Operations Research Letters}, 41(6):615--617, 2013.

\bibitem[Hoo91]{hoogeveen91}
J.~A. Hoogeveen.
\newblock Analysis of {C}hristofides' heuristic: Some paths are more difficult
  than cycles.
\newblock {\em Operations Research Letters}, 10:291--295, 1991.

\bibitem[Lov74]{Lov74}
L.~Lov\'{a}sz.
\newblock {L}ecture.
\newblock {C}onference of {G}raph {T}heory. {P}rague, 1974.

\bibitem[Lov79]{Lov79}
L.~Lov\'{a}sz.
\newblock {C}ombinatorial problems and exercises.
\newblock North-Holland, 1979.

\bibitem[Sch03]{Sch03}
A.~Schrijver.
\newblock {\em {C}ombinatorial {O}ptimization: {P}olyhedra and {E}fficiency,
  {A}lgorithms and {C}ombinatorics}, volume~24.
\newblock Springer, Berlin, 2003.

\bibitem[Seb13]{sebo13}
A.~Seb\H{o}.
\newblock Eight-fifth approximation for the path {TSP}.
\newblock In {\em Proceedings of the 16th Conference on Integer Programming and
  Combinatorial Optimization}, pages 362--374, 2013.

\bibitem[SV14]{SV12}
A.~Seb\H{o} and J.~Vygen.
\newblock Shorter tours by nicer ears: 7/5-approximation for the graph-{TSP},
  3/2 for the path version, and 4/3 for two-edge-connected subgraphs.
\newblock {\em Combinatorica}, pages 1--34, 2014.

\bibitem[Vyg12]{vygen13}
J.~Vygen.
\newblock {N}ew approximation algorithms for the {TSP}.
\newblock {\em Optima: Mathematical Optimization Society Newsletter}, 90(1-12),
  2012.

\end{thebibliography}

\end{document}